\documentclass[10pt]{amsart}

\usepackage[margin=2.625cm]{geometry}
\usepackage[numbers,sort&compress]{natbib}
\usepackage{booktabs,listings,graphicx,subfig}
\usepackage{amsaddr,amsmath,amssymb,eucal,nicefrac}
\usepackage[usenames,dvipsnames,svgnames,table]{xcolor}
\usepackage[heightadjust=all,valign=c]{floatrow}\usepackage{fr-subfig}
\usepackage[citecolor=Blue,colorlinks=true,linkcolor=Blue,urlcolor=Blue]{hyperref}

\usepackage{tikz,pgfplots}
\usetikzlibrary{decorations.markings,patterns}\pgfplotsset{compat=newest,grid style=dashed}
\tikzset{->-/.style={decoration={markings,mark=at position 0.5 with {\arrow{>}}},postaction=decorate}}
\tikzset{/pgf/number format/.cd,set thousands separator={{$^{\prime}$}},set decimal separator={{.}},precision=3,fixed}

\numberwithin{equation}{section}
\numberwithin{equation}{subsection}

\newcommand{\doi}[1]{DOI \href{http://dx.doi.org/#1}{\texttt{#1}}}

\begin{document}

\lstset{language=C++,tabsize=2,mathescape,escapechar={\$},basicstyle=\footnotesize,commentstyle=\ttfamily,keywordstyle=\bfseries,stringstyle=\ttfamily,morekeywords={from,then,to},numbers=left,numberstyle=\tiny,numbersep=-8pt,firstnumber=1}

\title{Time parallel gravitational collapse simulation}

\author{Andreas Kreienbuehl}
\address{Center for Computational Sciences and Engineering, Lawrence Berkeley National Laboratory, 1 Cyclotron Road, Berkeley CA 94720, United States of America}
\curraddr{}
\email{akreienbuehl@lbl.gov}
\thanks{}

\author{Pietro Benedusi}
\address{Institute of Computational Science, Faculty of Informatics, Universit{\`a} della Svizzera italiana, Via Giuseppe Buffi 13, 6904 Lugano, Switzerland}
\curraddr{}
\email{pietro.benedusi@usi.ch}
\thanks{}

\author{Daniel Ruprecht}
\address{School of Mechanical Engineering, University of Leeds, Woodhouse Lane, Leeds LS2 9JT, United Kingdom}
\curraddr{}
\email{d.ruprecht@leeds.ac.uk}
\thanks{}

\author{Rolf Krause}
\address{Institute of Computational Science, Faculty of Informatics, Universit{\`a} della Svizzera italiana, Via Giuseppe Buffi 13, 6904 Lugano, Switzerland}
\curraddr{}
\email{rolf.krause@usi.ch}
\thanks{}

\subjclass[2010]{35Q76, 65M25, 65Y05, 83C57}

\keywords{Einstein-Klein-Gordon gravitational collapse, Choptuik scaling, Parareal, spatial coarsening, load balancing, speedup}

\date{\today}

\dedicatory{}

\begin{abstract}
This article demonstrates the applicability of the parallel-in-time method Parareal to the numerical solution of the Einstein gravity equations for the spherical collapse of a massless scalar field. To account for the shrinking of the spatial domain in time, a tailored load balancing scheme is proposed and compared to load balancing based on number of time steps alone. The performance of Parareal is studied for both the sub-critical and black hole case; our experiments show that Parareal generates substantial speedup and, in the super-critical regime, can reproduce Choptuik's black hole mass scaling law.
\end{abstract}

\maketitle

\section{Introduction}

Einstein's field equations of general relativity (GR) consist of ten coupled, non-linear, hyperbolic-elliptic partial differential equations (PDEs). Because gravity couples to all forms of energy, there is an enormous dynamic range of spatiotemporal scales in GR. Hence, usually only the application of advanced numerical methods can provide solutions and in numerical relativity~\cite{Alcubierre2008a,BaumgarteAndShapiro2010a} extensive use of high-performance computing (HPC) is made~\cite{LoefflerEtAl2012a,KidderEtAl2000a}.

Today, almost all HPC architectures are massively parallel systems connecting large numbers of compute nodes by a high-speed interconnect. In numerical simulations, the power of these systems can only be harnessed by algorithms that feature a high degree of concurrency; every algorithm with strong serial dependencies can only provide inferior performance on massively parallel computers. For the solution of PDEs, parallelization strategies have been developed mainly for spatial solvers. However, in light of the rapid increase in the number of cores in supercomputers, methods that offer additional concurrency along the temporal axis have recently begun to receive more attention.

The idea of parallelization-in-time was introduced in $1964$~\cite{Nievergelt1964}. In the $1980$s and $1990$s, time and spacetime multigrid methods were studied~\cite{Hackbusch1984,Horton1992,HortonEtAl1995}. More recently, the now widely used time parallel method Parareal was proposed~\cite{LionsEtAl2001}. Other recently introduced parallel-in-time methods are PFASST~\cite{Minion2010,EmmettMinion2012}, RIDC~\cite{ChristliebEtAl2010}, or MGRIT~\cite{FalgoutEtAl2014_MGRIT}. A historical overview is offered in~\cite{Gander2015_Review}.

Given the demonstrated potential of parallel-in-time integration methods for large-scale parallel simulations~\cite{SpeckEtAl2012}, these methods could be beneficial for the numerical relativity community. However, their application is not straightforward and often it is unclear \textit{a priori} if good performance can be achieved. In this article, we therefore investigate the \textit{principal applicability} of the time parallel Parareal method to solving Einstein's equations describing spherical, gravitational collapse of a massless scalar field. The system is also referred to as an Einstein-Klein-Gordon system because it is equivalent to a Klein-Gordon equation expressed in the context of GR, \textit{i.e.}\ on a back-reacting, curved geometry. It defines a basic gravitational field theory and is of interest therefore not only in numerical relativity but also in, \textit{e.g.}, quantum gravity~\cite{Husain2009a,ZiprickAndKunstatter2009a,KreienbuehlEtAl2012a}. A summary of numerically derived results is given in~\cite{GundlachAndMartinGarcia2007a}; the work by Choptuik~\cite{Choptuik1993a} brought forward novel, physical results and is of particular interest here because we will show that Parareal correctly reproduces the expected mass scaling law.

Mathematical theory shows that Parareal performs well for diffusive problems with constant coefficients~\cite{GanderVandewalle2007_SISC}. For diffusive problems with space- or time-dependent coefficients, numerical experiments show that Parareal can converge quickly too~\cite{KreienbuehlEtAl2015}. However, given the theory for basic constant-coefficient hyperbolic PDEs~\cite{GanderVandewalle2007_SISC}, it can be expected that Parareal applied to convection dominated problems converges too slowly for meaningful speedup to be possible. Special cases with reasonable performance are discussed in~\cite{Gander2008} and for certain hyperbolic PDEs it was found that some form of stabilization is required for Parareal to provide speedup~\cite{GanderPetcu2008,RuprechtKrause2012,DaiEtAl2013,ChenEtAl2014}. Surprisingly, no stabilization is required for the equations describing gravitational collapse; we demonstrate that plain Parareal can achieve significant speedup. A detailed analytical investigation of why this is the case would definitely be of interest but is left out for future work. One reason could be that we solve in characteristic coordinates for which the discretization is aligned with the directions of propagation~\cite{Gander2008a,Kreienbuehl2011a}.

The article is structured as follows: In Section~\ref{s:equations} we define the system of Einstein field equations that we solve using Parareal. In addition, we give details on the numerical approach and discuss the interplay between Parareal and the particular structure of the spatial mesh. In Section~\ref{s:parareal} we discuss the Parareal method. Then, in Section~\ref{s:results} numerical results are presented. Finally, in Section~\ref{s:conclusion} we conclude with a summary and discussion.

\section{\label{s:equations}Equations}

\subsection{Gravitational collapse}

The Einstein field equations in Planck units normalized to $4\pi G/c^4=1$ are
\begin{equation}\label{e:einstein_equations}
	G_{\mu\nu}=2T_{\mu\nu},
\end{equation}
where $\mu,\nu\in\{0,1,2,3\}$ index time (via $0$) and space (via $1$, $2$, and $3$).\footnote{We omit the addition of the cosmological constant term $\Lambda g_{\mu\nu}$ on the left-hand side in Equation~\eqref{e:einstein_equations} because observations suggest $0<\Lambda\ll1$ (see, \textit{e.g.},~\cite{KomatsuEtAl2009a}); the term's impact on black hole formation as studied here can be neglected.} Once the non-gravitational matter content is specified by a definition of the energy-momentum tensor $T_{\mu\nu}$, possibly along with equations of state that together satisfy the continuity equations $\nabla^{\mu}T_{\mu\nu}=0$, Equation~\eqref{e:einstein_equations} defines a set of ten partial differential equations for ten unknown metric tensor field components $g_{\mu\nu}$.\footnote{We use the Einstein summation convention.} In all generality, the equations are coupled, non-linear, and hyperbolic-elliptic in nature. Six of the ten equations are hyperbolic evolution equations, while the remaining four are elliptic constraints on the initial data; they represent the freedom to choose spacetime coordinates. For the matter content, we consider a minimally coupled massless scalar field $\phi$ with energy-momentum tensor
\begin{equation}
	T_{\mu\nu}=\nabla_{\mu}\phi\nabla_{\nu}\phi-\frac{1}{2}g_{\mu\nu}g^{\alpha\beta}\nabla_{\alpha}\phi\nabla_{\beta}\phi.
\end{equation}
For the metric tensor field $g_{\mu\nu}$ in spherical symmetry it is natural to introduce a para\-metri\-zation in terms of Schwarz\-schild coordinates $(t,r)$. Here, $t$ is the time coordinate of a stationary observer at infinite radius $r$, which measures the size of spheres centered at $r=0$. In~\cite{Choptuik1993a} the resulting Einstein field equations are analyzed numerically. In particular, adaptive mesh refinement~\cite{BergerAndOliger1984a} is used to resolve the black hole formation physics. In~\cite{Garfinkle1994a} the same investigation is carried out in double null or characteristic coordinates $(\tau,\rho)$ without mesh refinement (see, however,~\cite{PretoriusAndLehner2004a,Thornburg2011a}). Finally, in~\cite{KreienbuehlEtAl2012a} the effect of quantum gravity modifications on the collapse is studied in adjusted characteristic coordinates. Here we use characteristic coordinates $(\tau,\rho)$ as well but exclude quantum gravity modifications. Also, for simplicity, we will refer to $\tau$ as a time coordinate and to $\rho$ as a space coordinate.

Making the \textit{ansatz}
\begin{equation}
	g_{\mu\nu}\text{d}x^{\mu}\text{d}x^{\nu}=-2\partial_{\rho}rH\text{d}\tau\text{d}\rho+r^2(\text{d}\vartheta^2+[\sin(\vartheta)\text{d}\varphi]^2)
\end{equation}
for the metric tensor field and using an auxiliary field $h$ for the spacetime geometry along with an auxiliary field $\Phi$ for the matter content, the complete field equations are
\begin{equation}\label{e:temporal_dynamics}
	\partial_{\tau}r=-\frac{h}{2},\qquad\partial_{\tau}\Phi=\frac{(H-h)(\Phi-\phi)}{2r},
\end{equation}
for $r$ and $\Phi$, and
\begin{equation}\label{e:spatial_dynamics}
	\partial_{\rho}\phi=\frac{\partial_{\rho}r}{r}(\Phi-\phi),\qquad\partial_{\rho}H=\frac{\partial_{\rho}r}{r}H(\Phi-\phi)^2,\qquad\partial_{\rho}h=\frac{\partial_{\rho}r}{r}(H-h),
\end{equation}
for $\phi$, $H$, and $h$ (see~\cite{Garfinkle1994a}). Overall the system can be seen as a wave equation for the massless scalar field $\phi$ on a back-reacting, curved geometry. Boundary conditions at $(\tau,\rho=\tau)$ are $r=0$ and regularity of $\Phi$, $\phi$, $H$, and $h$, which implies $\Phi=\phi$ and $H=h$ at the boundary~\cite{Christodoulou1993a,Kreienbuehl2011a}. Consistent initial data at $(\tau=0,\rho)$ are
\begin{equation}
	r=\frac{\rho}{2},\qquad\Phi=(1+\rho\partial_{\rho})\phi,
\end{equation}
where we choose for $\phi$ the Gaussian wave packet
\begin{equation}\label{e:initial_gaussian}
	\phi(0,\rho)=\phi_0\frac{\rho^3}{1+\rho^3}\exp\left(-\left[\frac{\rho-\rho_0}{\delta_0}\right]^2\right).
\end{equation}
We also performed tests for initial data similar in shape to the hyperbolic tangent function much like Choptuik did in~\cite{Choptuik1993a} for purely serial time stepping. Since in this case we found Parareal's performance to resemble strongly that for the case of the Gaussian wave packet we do not include these results here. The initial scalar field configuration is thus characterized by an amplitude $\phi_0$, mean position $\rho_0$, and width $\delta_0$. Depending on the value of these parameters, the solution to Equations~\eqref{e:temporal_dynamics} and~\eqref{e:spatial_dynamics} can describe a bounce of the wave packet or black hole formation near the boundary at $r=0$. A black hole appears when the outward null expansion
\begin{equation}\label{e:outward_expansion}
	\Theta^+=\frac{1}{r}\sqrt{\frac{2h}{H}},
\end{equation}
which measures the relative rate of change of a cross-sectional area element of a congruence of out-going null curves, approaches zero~\cite{Poisson2004a}. The black hole mass is
\begin{equation}\label{e:black_mass}
	M=\frac{r}{2},
\end{equation}
evaluated at the point $(\tau^+,\rho^+)$ toward which $\Theta^+$ vanishes.

\subsection{\label{s:numerical_solution}Numerical solution}

\begin{figure}[htbp]
	\captionsetup[subfigure]{justification=centering}
	\centering
	\ffigbox{
		\begin{subfloatrow}[2]
			\ffigbox[\FBwidth]{\caption{\label{f:generic_grid}The numerical domain. It is parametrized by the characteristic coordinates $\tau$ and $\rho$.}}{
				\includegraphics{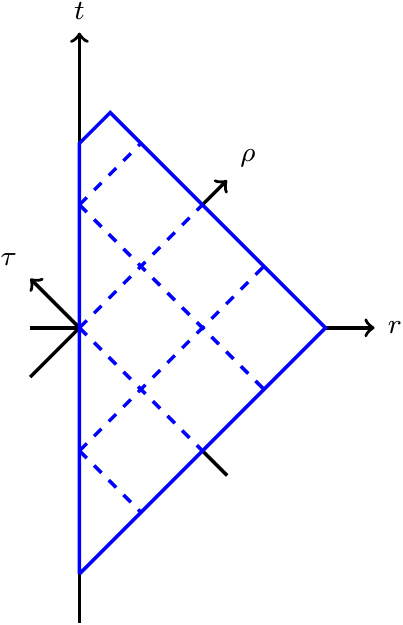}}
			\ffigbox[\FBwidth]{\caption{\label{f:snapshots}Scalar field solution snapshots for a black hole-free setting. The peak of the Gaussian evolves along the constant coordinate value $\rho\approx20$, which is also when the bounce occurs in $\tau$.}}{
				\includegraphics{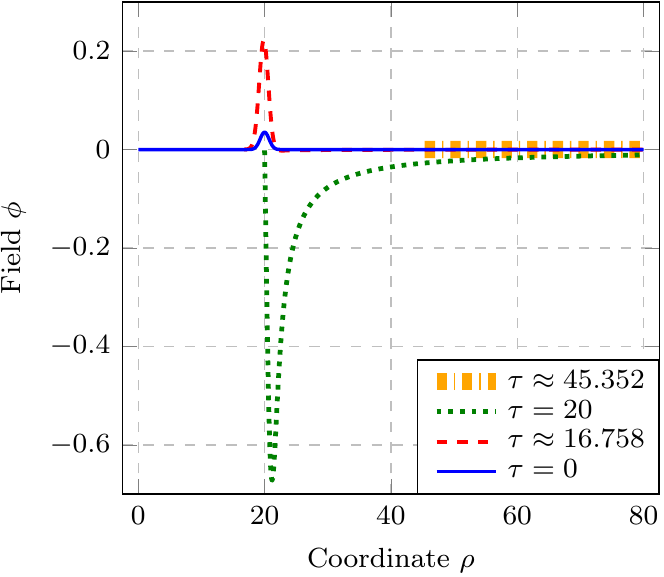}}
		\end{subfloatrow}
	}{\caption{\label{f:subcritical_case}The computational domain (left) and sub-critical gravitational scalar field evolution (right).}}
\end{figure}

The numerical grid is depicted in Figure~\ref{f:generic_grid}. It is parametrized by the characteristic coordinates $\tau$ and $\rho$, which are used for numerical integration; $\tau$ is used as coordinate representing time and $\rho$ as coordinate representing space. Integration thus takes place on a right triangle with initial data defined along the lower right-hand leg. Clearly, the spatial domain becomes smaller as the solution is advanced in $\tau$. Note that the domain is not exactly a right triangle because at the upper-most corner a small sub-triangle is missing. This ``buffer'' zone of extent $\lambda$ is needed for the spatial part of the numerical stencil to fit. The computational domain thus consists of all points $(\tau,\rho)\in[0,L-\lambda]\times[0,L]$ with $L=80$, $\lambda=0.625$, and $\rho\geq\tau$.

\thisfloatsetup{subfloatrowsep=quad,valign=b}

As a time stepping method for the solution of the equations in~\eqref{e:temporal_dynamics}, we use a second-order Lax-Wendroff Richtmyer two-step method on a fine spacetime grid (see~\cite{Kreienbuehl2011a}).
To employ the time parallel method Parareal (see Section~\ref{s:parareal}), we need a second, computationally cheap, time integration method.
Here, we choose the explicit first-order Euler method on a coarse spacetime mesh. For Parareal to be efficient, the cost of the coarse method has to be small compared to that of the fine one: by choosing a simple first-order method on the coarse grid for $\mathcal{C}$ we obtain a good coarse-to-fine ratio (see Section~\ref{s:speedup}). For optimal speedup, the right balance between the difference in accuracy and difference in cost between $\mathcal{C}$ and $\mathcal{F}$ has to be found.

For the integration in space of the equations in~\eqref{e:spatial_dynamics} we use a second-order Runge-Kutta method~\cite{Kreienbuehl2011a}. Snapshots of scalar field evolution resulting from the chosen fine grid discretization are shown in Figure~\ref{f:snapshots}, where $\phi$ evolves along constant lines of $\rho$ until a bounce occurs at $r=0$. The figure also shows how the size of the domain decreases during the evolution: for $\tau=0$ the left boundary is at $\rho=0$ while for $\tau=20$ it is at $\rho=20$.

\subsection{\label{s:mass_scaling}Mass scaling}

\begin{figure}[htbp]
	\captionsetup[subfigure]{justification=centering}
	\centering
	\ffigbox{
		\begin{subfloatrow}[2]
			\ffigbox[\FBwidth]{\caption{\label{f:collapse_grid}The simulation terminates at $\tau^+$, when a black hole forms at $\rho^+$.}}{
				\includegraphics{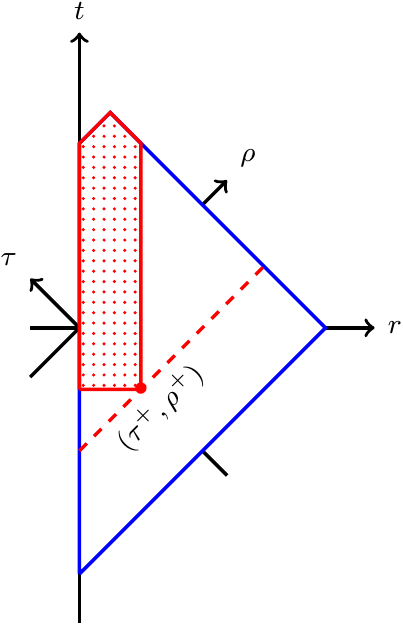}}
			\ffigbox[\FBwidth]{\caption{\label{f:expansion_overview}Minimal weighted outward null expansion indicating a bounce (top) and black hole formation (bottom) are shown.}}{
				\includegraphics{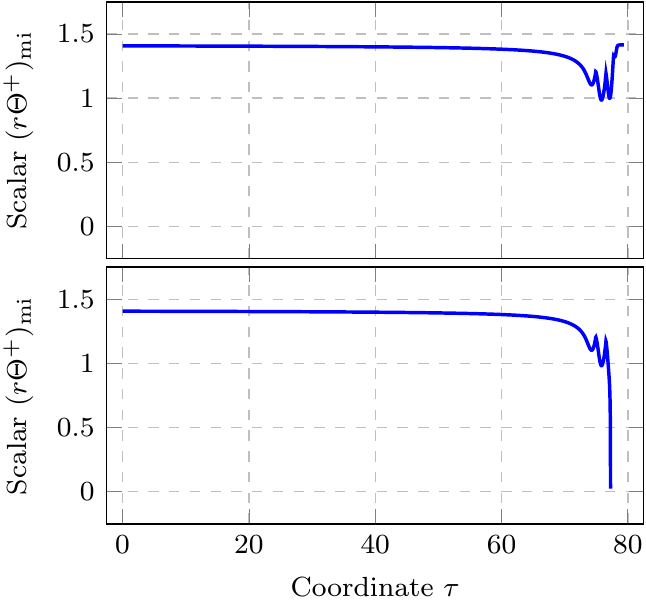}}
		\end{subfloatrow}
	}{\caption{\label{f:supercritical_case}Illustrations to clarify super-critical gravitational collapse.}}
\end{figure}

In practice, the simulation terminates when a black hole forms because $H$ grows without bound in this case (see~\cite{Christodoulou1993a} for details). Figure~\ref{f:collapse_grid} provides a simplified illustration of a black hole region (dotted portion) and shows where the simulation comes to a halt (dashed line). Thus, to determine the black hole mass $M$, we record minimal expansion values via the scalar $(r\Theta^+)_{\text{mi}}=\min_{\rho}\{r\Theta^+\}$ derived from Equation~\eqref{e:outward_expansion}. The last such recorded minimal value before the termination of the simulation defines a characteristic coordinate $(\tau^+,\rho^+)$ (see again Figure~\ref{f:collapse_grid}), which we can use to define an $r$ and $M$ via Equation~\eqref{e:black_mass}. The scalar $(r\Theta^+)_{\text{mi}}$ approaches $0$ when $(\tau,\rho)$ nears $(\tau^+,\rho^+)$, as is shown in the lower portion of Figure~\ref{f:expansion_overview}.

Based on numerical experiments, Choptuik presents, among other things, a relation between the amplitude $\phi_0$ of the Gaussian in Equation~\eqref{e:initial_gaussian} and the black hole mass $M$~\cite{Choptuik1993a}. He shows that there is a critical value $\phi^{\star}_0$ such that for $\phi_0<\phi^{\star}_0$ there is a bounce (sub-critical case), while for $\phi_0>\phi^{\star}_0$ there is a black hole (super-critical case). Based thereon, he demonstrates that the black hole mass scales with $\phi_0-\phi^{\star}_0>0$ according to the law $M\propto(\phi_0-\phi^{\star}_0)^{\gamma}$ with $\gamma$ being a positive constant of the same value for various initial data profiles. We demonstrate that Parareal can correctly capture this black hole mass scaling law although our coarse level Euler method alone cannot. Also, Parareal requires less wall-clock time than $\mathcal{F}$, which can be beneficial for the investigation of the high-accuracy demanding critical solution~\cite{Choptuik1993a,GundlachAndMartinGarcia2007a} that requires the simulation of numerous black holes~\cite{Garfinkle1994a}. This analysis however is omitted in this article and left for future work.

\thisfloatsetup{subfloatrowsep=quad,valign=b}

\section{\label{s:parareal}Parareal}

\subsection{Algorithm}

Parareal~\cite{LionsEtAl2001} is a method for the solution of initial value problems
\begin{equation}
	\partial_{\tau}u(\tau)=f(\tau,u(\tau)),\qquad u(0)=u_0,\qquad0\leq\tau\leq T.
\end{equation}
Here, as is outlined in the previous section, $f$ comes from discretizing Equations~\eqref{e:temporal_dynamics} and \eqref{e:spatial_dynamics}, and $T=L-\lambda$ marks the end time. Parareal starts with a decomposition of the time domain into $N_{\text{pr}}$ temporal subintervals (TSs) defined in terms of times $\tau^p$ such that
\begin{equation}
	[\tau^1,\tau^2]\cup...\cup[\tau^{N_{\text{pr}}-1},\tau^{N_{\text{pr}}}]=[0,L-\lambda].
\end{equation}
Now denote by $\mathcal{F}$ some serial time integration method of high accuracy and cost (in our case this is the second-order Lax-Wendroff Richtmyer two-step method), and by $\mathcal{C}$ a cheap and possibly much less accurate method (in our case this is the explicit first-order Euler method). Instead of running the fine method subinterval by subinterval serially in time, Parareal performs the iteration
\begin{equation}\label{e:parareal_algorithm}
	u^{p+1}_{[i+1]}=\mathcal{C}\left(u^p_{[i+1]}\right)-\mathcal{C}\left(u^p_{[i]}\right)+\mathcal{F}\left(u^p_{[i]}\right),
\end{equation}
where super-scripts index time or process number $p\in\{1,...,N_{\text{pr}}\}$ and sub-scripts iterations $i\in\{1,...,N_{\text{it}}\}$. The advantage is that the expensive computation of the fine method can be performed in parallel over all TSs at once. Here, we assume that the number of TSs is equal to the number $N_{\text{pr}}$ of cores (or \textit{processes}) used for the time direction. Good speedup can be obtained if $\mathcal{C}$ is fast in comparison to $\mathcal{F}$ but still accurate enough for Parareal to converge rapidly. See Section~\ref{s:speedup} for a more detailed discussion of Parareal's speedup.

In Section~\ref{s:numerical_solution} we hinted at the interchangeability of the characteristic coordinates $\tau$ and $\rho$ for the numerical integration. Therefore, theoretically, Parareal could \textit{also} be used for the spatial integration to \textit{simultaneously} parallelize both time and space. However, such an interweaving of two Parareal iterations is not discussed in this article; it is put aside for future work.

\subsection{Spatial coarsening in Parareal}

In order to make $\mathcal{C}$ cheaper and improve speedup, we not only use a less accurate time stepper for $\mathcal{C}$ but also employ a coarsened spatial discretization with a reduced number of degrees-of-freedom. Therefore, we need a spatial interpolation $\mathbf{I}$ and restriction $\mathbf{R}$ operator. In this case (see, \textit{e.g.},~\cite{FischerEtAl2005}), the Parareal algorithm is given by
\begin{equation}
	u^{p+1}_{[i+1]}=\mathbf{I}\mathcal{C}\left(\mathbf{R}u^p_{[i+1]}\right)-\mathbf{I}\mathcal{C}\left(\mathbf{R}u^p_{[i]}\right)+\mathcal{F}\left(u^p_{[i]}\right).
\end{equation}
As restriction operator $\mathbf{R}$ we use point injection. For the interpolation operator $\mathbf{I}$ we use polynomial (\textit{i.e.}\ Lagrangian) interpolation of order $3$, $5$, and $7$.\footnote{We also tested barycentric interpolation~\cite{BerrutAndTrefethen2004a,FloaterAndHormann2007a} but found the performance in terms of runtimes and speedup (see Sections~\ref{s:speedup} and~\ref{s:results}) to be inferior.} It has been shown that, even for simple toy problems, convergence of Parareal can deteriorate if spatial coarsening with low-order interpolation is used. As demonstrated in Section~\ref{s:sub_critical}, this also holds true for the here studied problem.

\subsection{\label{s:implementation}Implementation}

We have implemented two different realizations of Parareal. In a ``standard'' version $\mathcal{P}_{\text{st}}$ (see Listing~\ref{l:standard_implementation}), the Parareal correction is computed on each TS up to a uniformly prescribed iteration number. In contrast, in the ``modified'' implementation $\mathcal{P}_{\text{mo}}$ (see Listing~\ref{l:modified_implementation}), Parareal corrections are only performed on TSs where the solution may not yet have converged. Because Parareal always converges at a rate of at least one TS per iteration, we only iterate on a TS if its assigned \texttt{MPI} rank is greater than or equal to the current Parareal iteration number (see line $8$ in Listing~\ref{l:modified_implementation}). Otherwise, no further iterations are needed and performed, and the process remains idle. Thus, as the iteration progresses, more and more processes enter an idle state.
In an implementation to be realized in future work, the criterion for convergence used here will be replaced by a check for some residual tolerance~\cite{Aubanel2011a}. This could negatively affect the observed performance since it requires essentially one more iteration to compute the residual.\footnote{In~\cite{Aubanel2011a} a version of Parareal is discussed that can be used to proceed the integration beyond a given end time. It is based on an optimized scheduling of those tasks which become idle in our implementation.}
It also bears mentioning that it has very recently been demonstrated that parallel-in-time integration methods are good candidates to provide algorithm-based fault tolerance~\cite{NielsenHesthaven2016,SpeckRuprecht2016}.

Another difference between the standard and modified implementation is that in the former, after each time parallel fine evolution, a copy of the fine grid solution has to be created (see line $10$ in Listing~\ref{l:standard_implementation}). In the modified Listing~\ref{l:modified_implementation} this copying is circumvented by the use of two alternating indices ``j'' and ``k'' in lines $9$ and $10$, respectively. The iteration number determines their value which, in turn, determines the fine grid solution buffer that is used to send or receive data by means of the corresponding \texttt{MPI} routines (see lines $14$ and $22$ in Listing~\ref{l:modified_implementation}). The two implementations also have slightly different requirements in terms of storage. As can be seen in line $15$ in Listing~\ref{l:standard_implementation}, in $\mathcal{P}_{\text{st}}$ on the first TS or, equivalently, for the first \texttt{MPI} rank, the fine grid solution has to be assigned initial data at the beginning of each iteration. This requires one additional buffer to be held in storage. Other than that both implementations need one coarse grid solution buffer and three fine grid buffers for each TS.

\newsavebox{\lsta}
\begin{lrbox}{\lsta}
	\begin{minipage}{0.45\textwidth}
		\begin{lstlisting}[frame=single]
		if $p$ > 1 then // Initialization
			Coarse(co; $\tau^1\to\tau^p$)
		Interp(co $\mapsto$ fi[0])
		if $p$ < $N_{\text{pr}}$ then // Prediction
			Coarse(co; $\tau^p\to\tau^{p+1}$)
		Interp(co $\mapsto$ fi[2])
		for i = 1 : $N_{\text{it}}$ do // Iteration
			if $p$ < $N_{\text{pr}}$ then
				Fine(fi[0]; $\tau^p\to\tau^{p+1}$)
			fi[1] = fi[0]
			fi[1] -= fi[2]
			if $p$ > 1 then
				MPI_Recv(fi[0]; $p\Leftarrow p-1$)
			else
				Init(fi[0])
			Restrict(fi[0] $\mapsto$ co)
			if $p$ < $N_{\text{pr}}$ then
				Coarse(co; $\tau^p\to\tau^{p+1}$)
			Interp(co $\mapsto$ fi[2])
			fi[1] += fi[2]
			if $p$ < $N_{\text{pr}}$ then
				MPI_Send(fi[1]; $p\Rightarrow p+1$)
		\end{lstlisting}
	\end{minipage}
\end{lrbox}

\newsavebox{\lstb}
\begin{lrbox}{\lstb}
	\begin{minipage}{0.45\textwidth}
		\begin{lstlisting}[frame=single]
		if $p$ > 1 then // Initialization
			Coarse(co; $\tau^1\to\tau^p$)
		Interp(co $\mapsto$ fi[0])
		if $p$ < $N_{\text{pr}}$ then // Prediction
			Coarse(co; $\tau^p\to\tau^{p+1}$)
		Interp(co $\mapsto$ fi[2])
		for i = 1 : $N_{\text{it}}$ do // Iteration
			if $p$ >= i then
				j = (i+1) $\%$ 2
				k = i $\%$ 2
				if $p$ < $N_{\text{pr}}$ then
					Fine(fi[j]; $\tau^p\to\tau^{p+1}$)
				if $p$ > i then
					MPI_Recv(fi[k]; $p\Leftarrow p-1$)
					fi[j] -= fi[2]
					Restrict(fi[k] $\mapsto$ co)
					if $p$ < $N_{\text{pr}}$ then
						Coarse(co; $\tau^p\to\tau^{p+1}$)
					Interp(co $\mapsto$ fi[2])
					fi[j] += fi[2]
				if $p$ < $N_{\text{pr}}$ then
					MPI_Send(fi[j]; $p\Rightarrow p+1$)
		\end{lstlisting}
	\end{minipage}
\end{lrbox}

\thisfloatsetup{subfloatrowsep=quad,valign=b}
\begin{figure}[htbp]
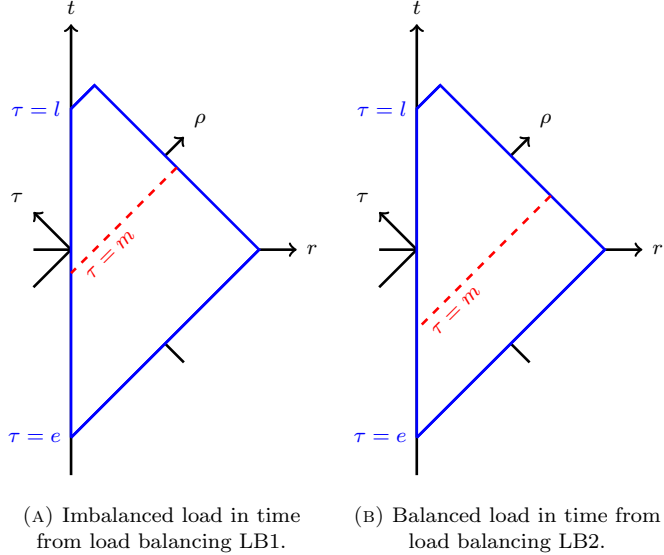

	\captionsetup[subfigure]{justification=centering}
	\centering
	\ffigbox{
		\begin{subfloatrow}[2]
			\ffigbox[\FBwidth]{\caption{\label{l:standard_implementation}The standard Parareal implementation $\mathcal{P}_{\text{st}}$.}}{
				\usebox{\lsta}}
			\ffigbox[\FBwidth]{\caption{\label{l:modified_implementation}The modified Parareal implementation $\mathcal{P}_{\text{mo}}$.}}{
				\usebox{\lstb}}
		\end{subfloatrow}\\
	}{\caption{\label{l:parareal_implementations}Pseudo code for the standard and modified Parareal implementation. The variable ``co'' denotes the coarse grid solution and ``fi'' an array of three fine grid buffers.}}
\end{figure}

\subsection{\label{s:speedup}Speedup}

We denote by $R_{\text{co}}$ the coarse and by $R_{\text{fi}}$ the fine time stepper's runtime. Recalling that $N_{\text{it}}$ denotes the number of iterations required for Parareal to converge given $N_{\text{pr}}$ processes, Parareal's theoretically achievable speedup is
\begin{equation}\label{e:speedup_formula}
	S=\left[\left(1+\frac{N_{\text{it}}}{N_{\text{pr}}}\right)\frac{R_{\text{co}}}{R_{\text{fi}}}+\frac{N_{\text{it}}}{N_{\text{pr}}}\right]^{-1}\leq\min\left\{\frac{N_{\text{pr}}}{N_{\text{it}}},\frac{R_{\text{fi}}}{R_{\text{co}}}\right\},
\end{equation}
as is discussed, \textit{e.g.}, in~\cite{Minion2010}. The estimate is valid only for the ideal case, where runtimes across subintervals are perfectly balanced. In the presence of load imbalances in time however, \textit{i.e.}\ differences in the runtimes of $\mathcal{C}$ and $\mathcal{F}$ across TSs, maximum speedup is reduced~\cite{KreienbuehlEtAl2015}. Because the spatial domain we consider is shrinking in time, a tailored decomposition of the time axis has to be used to provide well balanced computational load, as is discussed in the next section.

\subsection{Load balancing}\label{s:loadbalancing}

\begin{figure}[htbp]
	\captionsetup[subfigure]{justification=centering}
	\centering
	\ffigbox{
		\begin{subfloatrow}[2]
			\ffigbox[\FBwidth]{\caption{\label{f:lb1_grid}Imbalanced load in time from load balancing LB1.}}{
				\includegraphics{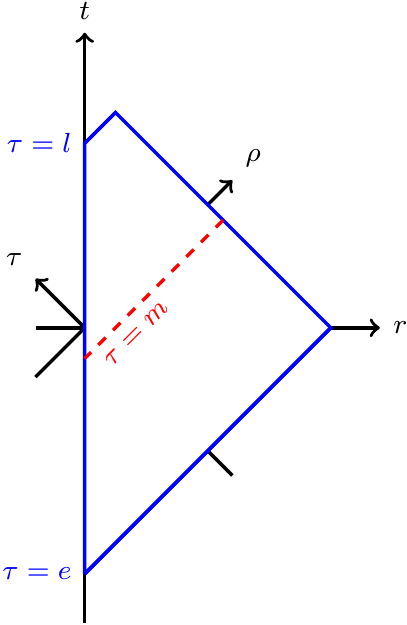}}
			\ffigbox[\FBwidth]{\caption{\label{f:lb2_grid}Balanced load in time from load balancing LB2.}}{
				\includegraphics{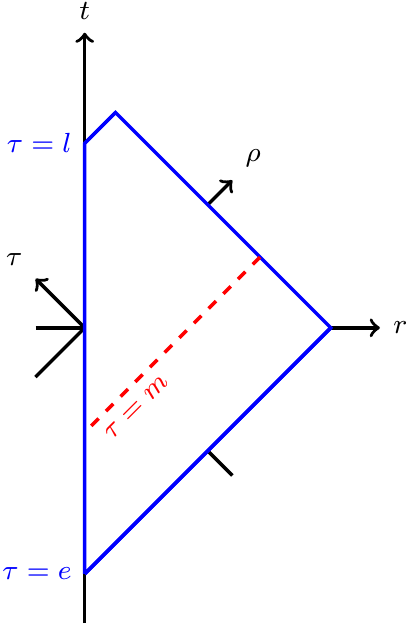}}
		\end{subfloatrow}
	}{\caption{\label{f:load_balancings}Illustration of two different approaches for the decomposition of the time domain, \textit{i.e.}\ LB1 (left) and LB2 (right).}}
\end{figure}

Because we integrate over a triangular computational spacetime domain (see Figure~\ref{f:generic_grid}), a straight forward, uniform partitioning of the time axis results in imbalanced computational load in time. The first load balancing (LB) strategy, which henceforth we will refer to as LB1, is based on this straight forward, basic decomposition of the time axis. It assigns to each TS the same \textit{number} of time steps without regard to their computational \textit{cost}. Because of the shrinking domain, TSs at later times carry fewer spatial degrees-of-freedom so that the per-process runtimes $R^p_{\text{co}}$ and $R^p_{\text{fi}}$ of the coarse and fine time stepper, respectively, are larger for the earlier TSs than for the later ones. Figure~\ref{f:lb1_grid} shows how this partition leads to an imbalanced computational load in time because the portion extending across the ``early-middle'' TS $[e,m]$ covers a larger area and thus a larger number of grid points than the portion over the ``middle-late'' TS $[m,l]$.

\thisfloatsetup{subfloatrowsep=quad,valign=b}

Figure~\ref{f:load_balancings} suggests that early in time TSs should have a shorter extent in time than later ones. Thus, in the second strategy, which in the following we will refer to as LB2, we also consider the \textit{cost} of time steps in order to balance the runtime $R^p_{\text{co}}+R^p_{\text{fi}}$ over all processes $p$. We use a decomposition of the time axis in TSs such that the sum of the total coarse and fine runtime is balanced over all TSs, \textit{i.e.}\ such that $R_{\text{co}}+R_{\text{fi}}=N_{\text{pr}}(R^p_{\text{co}}+R^p_{\text{fi}})$ for any process $p$. This is done by a bisection approach, making use of the fact the we use explicit rather than implicit time integrators (\textit{cf}.\ the discussion in~\cite{KreienbuehlEtAl2015}), and thus that the cost of a time step from $\tau$ to $\tau+\Delta\tau$ is directly proportional to the number of spatial degrees-of-freedom at $\tau$. Therefore, the total spacetime domain is first divided into two parts of roughly equal number of grid points as is sketched in Figure~\ref{f:lb2_grid}. Then, each part is divided again and again until the required number of TSs is reached. Note that this limits the possible numbers of TSs to powers of $2$.

\begin{figure}[htbp]
	\captionsetup[subfigure]{justification=centering}
	\centering
	\ffigbox{
		\begin{subfloatrow}[1]
			\ffigbox[\FBwidth]{\caption{\label{f:uneven_trace}\texttt{Vampir} trace for LB1. The Parareal runtime is $R_{\text{pa}}=7.964\ (\text{s})$.}}{
				\includegraphics[width=0.625\textwidth]{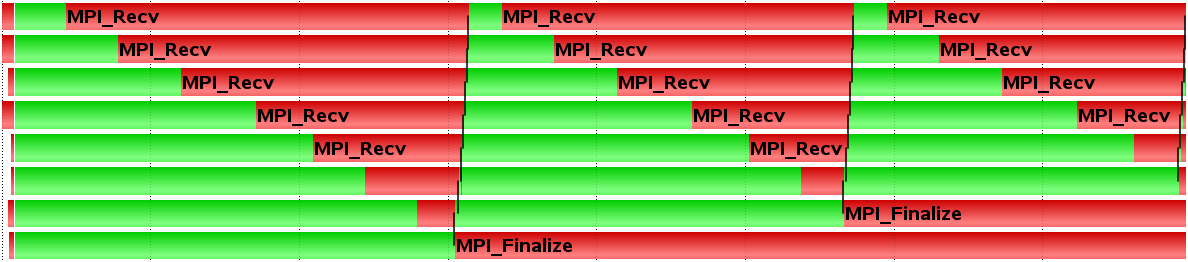}}
		\end{subfloatrow}\\
		\vspace{\baselineskip}
		\begin{subfloatrow}[1]
			\ffigbox[\FBwidth]{\caption{\label{f:even_trace}\texttt{Vampir} trace for LB2. The Parareal runtime is $R_{\text{pa}}=5.436\ (\text{s})$.}}{
				\includegraphics[width=0.625\textwidth]{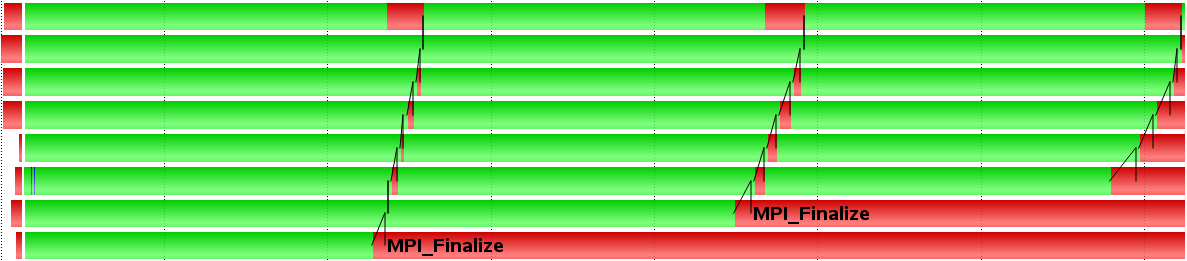}}
		\end{subfloatrow}
	}{\caption{\label{f:vampir_traces}\texttt{Vampir} traces for the implementation $\mathcal{P}_{\text{mo}}$ with $(N_{\text{pr}},N_{\text{it}})=(8,3)$ for two different load balancing strategies.}}
\end{figure}

Figure~\ref{f:vampir_traces} shows \texttt{Vampir}\footnote{\href{https://www.vampir.eu/}{https://www.vampir.eu/}} traces for one simulation featuring LB1 (Figure~\ref{f:uneven_trace}) and one LB2 (Figure~\ref{f:even_trace}). The horizontal axes correspond to runtime, while the vertical axes depict \texttt{MPI} rank numbers from $1$ (lower) to $8$ (upper). In each case, three Parareal iterations are performed. Green regions indicate the coarse and fine integrators carrying out work. Time spent in \texttt{MPI} receives (including waiting time) is shown in red. We observe how LB1 leads to load imbalance and incurs significant wait times in processes handling later TS. In contrast, the processes' idle times (shown in red) in \texttt{MPI} receives are almost invisible in the case of LB2. Elimination of wait times leads to a significant reduction in runtime and increase in speedup, as will be shown in Section~\ref{s:results}.

\thisfloatsetup{subfloatrowsep=quad,valign=b}

\section{\label{s:results}Results}

Speedup and runtime measurements were performed on the Cray XC40 supercomputer Piz Dora\footnote{\href{http://user.cscs.ch/computing\_systems/piz\_dora/}{http://user.cscs.ch/computing\_systems/piz\_dora/}} at the Swiss National Supercomputing Centre (CSCS) in Lugano, Switzerland. It features $1,\!256$ compute nodes, which all hold two $12$-core Intel Xeon E5-2690v3 processors. This results in a total of $30,\!144$ compute cores and a peak performance of $1.254$ PFlops; it occupies position $56$ in the Top500 November, $2014$ list.\footnote{\href{http://www.top500.org/list/2014/11}{http://www.top500.org/list/2014/11}} On Piz Dora, we used the GNU compiler collection\footnote{\href{https://gcc.gnu.org}{https://gcc.gnu.org}} version $4.9.2$ and the runtimes we provide do not include the cost of I/O operations.
Some simulations measuring convergence were performed on a machine located at the Universit{\`a} della Svizzera italiana that is maintained by members of the Institute of Computational Science of the Faculty of Informatics.\footnote{\href{https://www.ics.usi.ch/index.php/ics-research/resources}{https://www.ics.usi.ch/index.php/ics-research/resources}}

For the results presented in the following we use a coarse grid resolution of $(\Delta\tau)_{\text{co}}=(\Delta\rho)_{\text{co}}=\Delta_{\text{co}}=L/2,\!048\approx0.039$ and a fine grid resolution of $\Delta_{\text{fi}}=\Delta_{\text{co}}/8\approx0.005$. We have also determined a \textit{reference} solution to approximately measure the serial fine stepper's discretization error. For this we have used again the serial fine time stepper but with a step size of $\Delta_{\text{re}}=\Delta_{\text{fi}}/4\approx0.001$.

\subsection{Sub-critical}\label{s:sub_critical}

\begin{figure}[htbp]
	\captionsetup[subfigure]{justification=centering}
	\centering
	\ffigbox{
		\begin{subfloatrow}[2]
			\ffigbox[\FBwidth]{\caption{\label{f:early_conv}Parareal's defect over time for an early bounce scenario.}}{
				\includegraphics{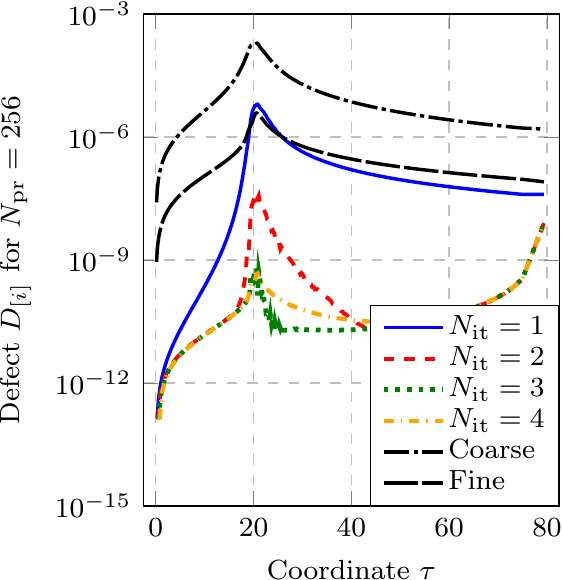}}
			\ffigbox[\FBwidth]{\caption{\label{f:late_conv}Defect of Parareal over time for a late bounce situation.}}{
				\includegraphics{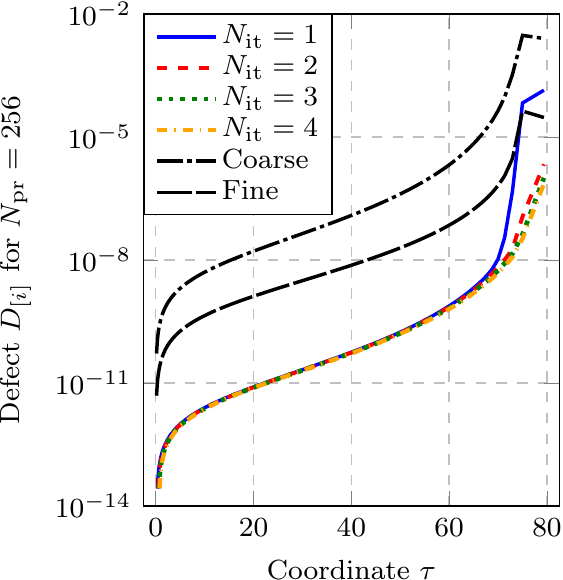}}
		\end{subfloatrow}
	}{\caption{\label{f:convergence_time}Defect in $r$ between Parareal and the fine method over time for fixed $N_{\text{pr}}=256$.}}
\end{figure}

First we consider the sub-critical case, where no black holes form. Figure~\ref{f:convergence_time} shows for $N_{\text{pr}}=256$ and two different sets of initial data parameters the relative defect
\begin{equation}
	D_{[i]}=\frac{\|r_{[i]}-r_{\text{fi}}\|_2}{\|r_{\text{fi}}\|_2},
\end{equation}
which measures the difference between the Parareal solution $r_{[i]}$ after $i$ iterations and the serial fine solution $r_{\text{fi}}$ as a function of the characteristic coordinate $\tau$. 

In Figure~\ref{f:early_conv} we use the initial data parameters $(\phi_0,\rho_0,\delta_0)=(0.035,20,1)$, which results in an ``early'' bounce of the wave packet at about $\tau=20$. For the simulations in Figure~\ref{f:late_conv}, the values are $(\phi_0,\rho_0,\delta_0)=(0.01,75,1)$, which leads to a ``late'' bounce at about $\tau=75$. Defects are plotted for $N_{\text{it}}\in\{1,2,3,4\}$ along with the serial coarse and fine solution's estimated discretization error $\|r_{\text{co}}-r_{\text{re}}\|_2/\|r_{\text{fi}}\|_2$ and $\|r_{\text{fi}}-r_{\text{re}}\|_2/\|r_{\text{fi}}\|_2$ labeled ``Coarse'' and ``Fine'', respectively. We observe that in Figure \ref{f:early_conv}, the data for $N_{\text{it}} = 3$ is somewhat jagged because for LB2 there are various start and end times of TSs near the bounce region. In any case, Parareal converges in two iterations: for $N_{\text{it}}=2$, the defect is below the discretization error for all $\tau$. In fact, without the bounce region near $\tau=20$, only one iteration would be required for convergence. For the late bounce scenario in Figure~\ref{f:late_conv}, we also observe that the rate of convergence at the final time $\tau=L-\lambda$ gives an indication of the convergence at all $\tau$. In the following we thus focus on convergence at the final time. Convergence for the other evolved field $\Phi$ is not shown but was found to be at least as good as for $r$.\footnote{Convergence seems to be unaffected by the load balancing. In tests not documented here we found that for LB1 it takes two iterations for Parareal to converge as well.}

\thisfloatsetup{subfloatrowsep=quad,valign=b}

\begin{figure}[htbp]
	\captionsetup[subfigure]{justification=centering}
	\centering
	\ffigbox{
		\begin{subfloatrow}[3]
			\ffigbox[\FBwidth]{\caption{\label{f:poly_conv}Defect for late bounce and interpolation order $3$.}}{
				\includegraphics{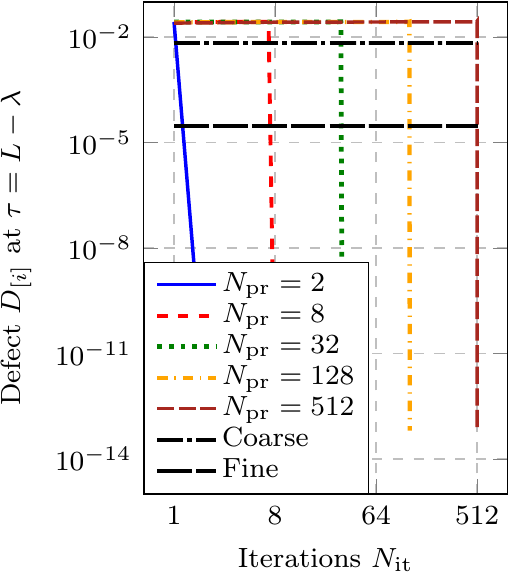}}
			\ffigbox[\FBwidth]{\caption{\label{f:sub_conv}Defect for late bounce and interpolation order $5$.}}{
				\includegraphics{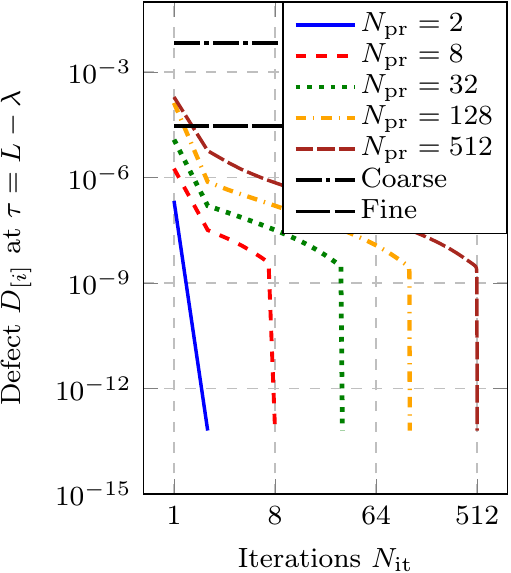}}
			\ffigbox[\FBwidth]{\caption{\label{f:sub_speed}Parareal speedup for fifth-order interpolation.}}{
				\includegraphics{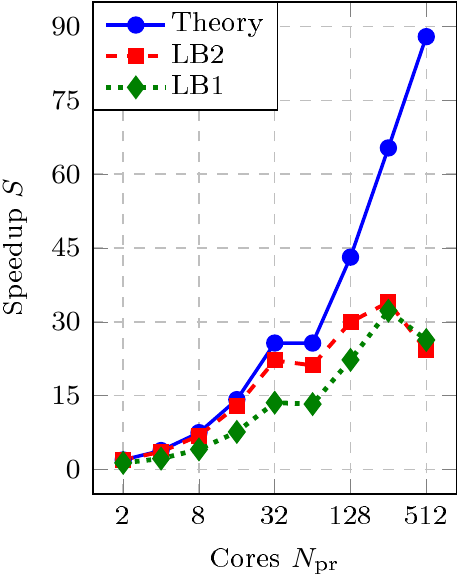}}
		\end{subfloatrow}
	}{\caption{\label{f:sub_performance}Parareal's performance for the sub-critical case in terms of convergence for polynomial interpolation orders $3$ and $5$, and in terms of speedup.}}
\end{figure}

Figures~\ref{f:poly_conv} and ~\ref{f:sub_conv} illustrate the defect of Parareal at the end of the simulation at $\tau=L-\lambda$ for various values of $N_{\text{pr}}$ with third-order interpolation (left) and fifth-order interpolation (right). For third-order interpolation, Parareal does not converge at all. The configuration stalls at a defect of about $10^{-2}$ until the iteration count equals $N_{\text{pr}}$. There, Parareal converges by definition but cannot provide any speedup. In contrast,  Parareal shows good convergence behavior for fifth-order interpolation. For $N_{\text{pr}}$ less than $64$, the defect of Parareal falls below the approximate discretization error of the fine method after a single iteration. Otherwise, for $N_{\text{pr}}\geq64$ up to $N_{\text{pr}}=512$, two iterations are required. 

The resulting speedups with correspondingly adjusted values for $N_{\text{it}}$ are shown in Figure~\ref{f:sub_speed} for both load balancing strategies (see the discussion in Section~\ref{s:loadbalancing}). In addition, the projected speedup according to Equation~\eqref{e:speedup_formula} is shown. 
The fine-to-coarse ratio $R_{\text{fi}}/R_{\text{co}}$ was determined experimentally and found to be about $74$.
Up to $N_{\text{pr}}=64$, for the advanced load balancing, speedup closely mirrors the theoretical curve while the basic load balancing performs significantly worse. For $N_{\text{pr}}\geq 64$, measured speedups fall short of the theoretical values, peak at $N_{\text{pr}}=256$, and then start to decrease.
Note that the theoretical model (blue line in Figure~\ref{f:sub_speed}) does take into account the scaling limit from the serial correction step according to Amdahl's law.
The difference between theory and measured speedup is therefore due to other overheads (communication and transfer between meshes) as analysed below.

Although the load balancing strategy LB2 results in significantly better speedup than the basic approach LB1, the peak value provided by both schemes is essentially the same. This is because for increasingly large numbers of cores, the computational load per TS eventually becomes small and imbalances in computational load insignificant. Instead, runtime is dominated by overhead from, \textit{e.g.}, communication in time. The communication load is independent of the chosen load balancing and depends solely on the number of TSs; for every TS one message has to be sent and received once per iteration (save for the first and last TS). Therefore, it can be expected that ultimately both approaches to load balancing lead to comparable peak values. Below we demonstrate that the saturation in speedup is related to a significant increase in time spent in \texttt{MPI} routines; eventually, communication cost starts to dominate over the computational cost left on each time slice and the time parallelization saturates just as spatial parallelization does.

\thisfloatsetup{subfloatrowsep=quad,valign=b}

\begin{figure}[htbp]
	\captionsetup[subfigure]{justification=centering}
	\centering
	\ffigbox{
		\begin{subfloatrow}[2]
			\ffigbox[\FBwidth]{\caption{\label{f:overhead_average}Average overhead.}}{
				\includegraphics{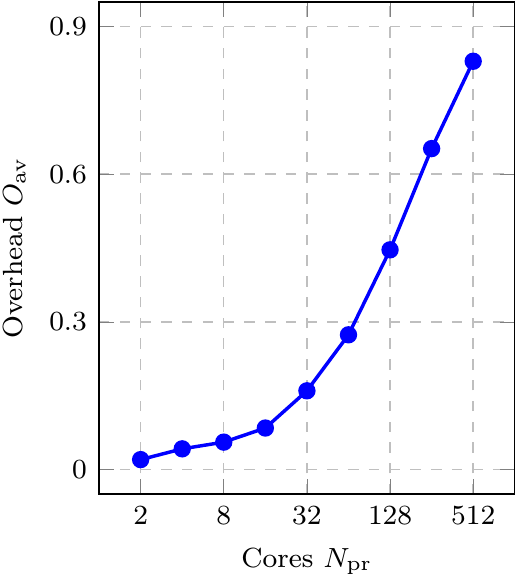}}
			\ffigbox[\FBwidth]{\caption{\label{f:overhead_stages}Overhead caused by three different Parareal stages.}}{
				\includegraphics{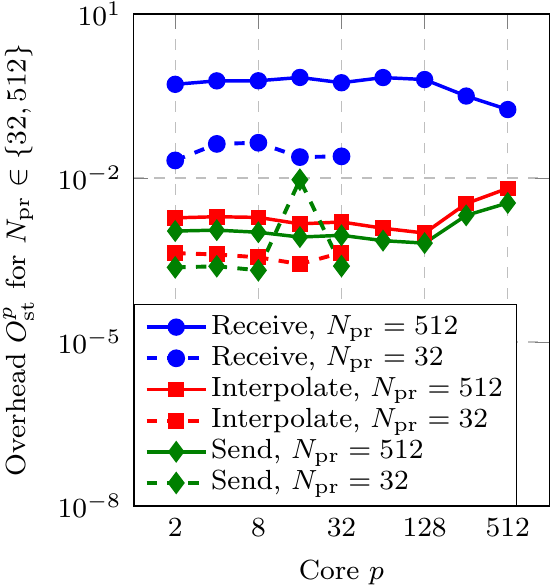}}
		\end{subfloatrow}
	}{\caption{\label{f:overhead_indicators}Overhead from communication and other sources increases with $N_{\text{pr}}$, which leads to Parareal's speedup decay.}}
\end{figure}

Figure~\ref{f:overhead_indicators} illustrates the reason behind the drop-off in speedup beyond $N_{\text{pr}}=256$. First, define
\begin{equation}\label{e:parareal_runtime}
	R^p_{\text{pa}}=R^p_{\text{co}}+R^p_{\text{fi}}+\sum_{\text{st}}R^p_{\text{st}},
\end{equation}
where $R^p_{\text{st}}$ denotes runtime spent in \textit{stages} that are \textit{different} from coarse and fine integration on the TS assigned to process $p$. For now, we consider only overhead from sending and receiving data as well as from interpolation; other overheads are not further analyzed here. Next, we introduce the \textit{total} overhead on a TS as the sum of all stage-runtimes or
\begin{equation}\label{e:overhead_total}
	O^p_{\text{to}}=\sum_{\text{st}}R^p_{\text{st}},
\end{equation}
which is also the runtime spent \textit{neither} in the coarse \textit{nor} fine integrator for a given $p$. The \textit{average} overhead is now defined as the geometric mean value of $O^p_{\text{to}}$ over all TSs, which is
\begin{equation}\label{e:overhead_average}
	O_{\text{av}}=\frac{\sum_{p=1}^{N_{\text{pr}}}O^p_{\text{to}}}{N_{\text{pr}}}.
\end{equation}
Finally, we define the relative overhead for individual \textit{stages} on a TS as
\begin{equation}\label{e:overhead_stages}
	O^p_{\text{st}}=\frac{R^p_{\text{st}}}{R^p_{\text{pa}}},
\end{equation}
where $R^p_{\text{pa}}$ is the runtime of Parareal at processor $p$. Ideally, as is assumed for the derivation of the speedup model given in Equation~\eqref{e:speedup_formula}, $R^p_{\text{co}}$ and $R^p_{\text{fi}}$ are the dominant costs. In this case, $R^p_{\text{co}}+R^p_{\text{fi}}\approx R^p_{\text{pa}}$ so that according to Equation~\eqref{e:parareal_runtime} we have $O^p_{\text{to}}\approx0$ and therefore $O_{\text{av}}\approx0$ by definition. However, as can be seen in Figure~\ref{f:overhead_average}, $O_{\text{av}}$ is small only for small values of $N_{\text{pr}}$. For $N_{\text{pr}}\geq32$ it increases rapidly, which indicates that the overhead from communication and other sources starts to play a more dominant role when $N_{\text{pr}}$ is increased.

Figure~\ref{f:overhead_stages} shows the relative overhead from Equation~\eqref{e:overhead_stages} for $N_{\text{pr}}\in\{32,512\}$ and $p\in\{1,...,N_{\text{pr}}\}$ for the three different stages $\text{st}\in\{\text{Interpolation},\text{Send},\text{Receive}\}$; ``Send'' and ``Receive'' refer to the corresponding \texttt{MPI} routines. There is a significant increase in relative overhead in all three stages as the number of cores grows, causing the eventual drop-off in speedup for increasing $N_{\text{pr}}$.

\thisfloatsetup{subfloatrowsep=quad,valign=b}

\subsection{Super-critical}

We consider now the more complex case in which a black hole forms at some time during the simulation. The goal is to compute the black hole's position via Equation~\eqref{e:outward_expansion} so that its mass can be determined from Equation~\eqref{e:black_mass} (see Section~\ref{s:mass_scaling}). Because the characteristic coordinates $(\tau,\rho)$ do not allow us to continue the simulation past the black hole formation event, we need a way to keep the simulation from terminating when $\Theta^+$ approaches $0$ (see Figure~\ref{f:expansion_overview}).

To avoid the need to adaptively modify the decomposition of the time domain, we carry out the super-critical case study using initial data parameter values near $(\phi_0,\rho_0,\delta_0)=(0.01,75,1)$, which we have also used for the results in Figure~\ref{f:late_conv}. With these parameters and in particular for $\phi_0\geq0.01$, for all investigated partitions of the time axis with $N_{\text{pr}}\leq256$, the black hole generated by the fine time integrator forms in the \textit{last} TS unless $\phi_0$ becomes too large ($\rho_0$ and $\delta_0$ are fix). Thus, Parareal can be used over all TSs except for the last one, where only the fine method is executed to compute the black hole's position. The C\texttt{++} implementation uses a \texttt{try}-\texttt{throw}-\texttt{catch} approach to prevent complete termination of the simulation; if the radicand in the definition of $\Theta^+$ in Equation~\eqref{e:outward_expansion} fails to be non-negative, an exception is thrown such that the Parareal iteration can continue. As the Parareal iteration converges and better and better starting values are provided for $\mathcal{F}$ on the last TS, the accuracy of the computed black hole position improves. A more general implementation aiming at production runs would need to allow for black hole formation in TSs before the last one but this is left for future work. In this article, the focus lies on investigating the \textit{principal applicability} of Parareal to the simulation of gravitational collapse. 

\thisfloatsetup{subfloatrowsep=quad,valign=b}
\begin{figure}[htbp]
	\captionsetup[subfigure]{justification=centering}
	\centering
	\ffigbox{
		\begin{subfloatrow}[2]
			\ffigbox[\FBwidth]{\caption{\label{f:super_conv}Choptuik scaling from Parareal.}}{
				\includegraphics{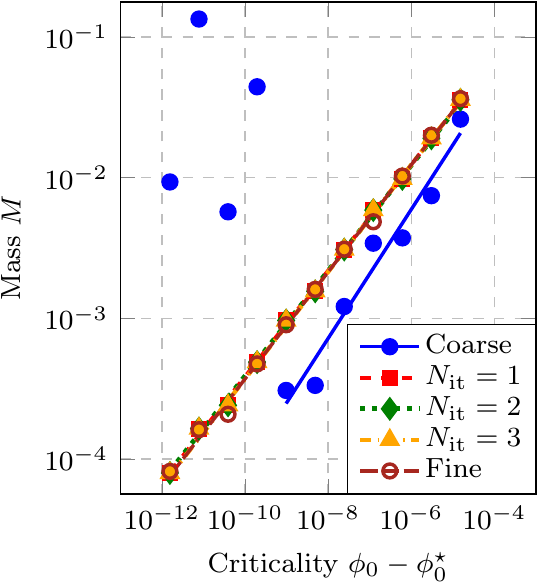}}
			\ffigbox[\FBwidth]{\caption{\label{f:super_speed}Parareal speedup.}}{
				\includegraphics{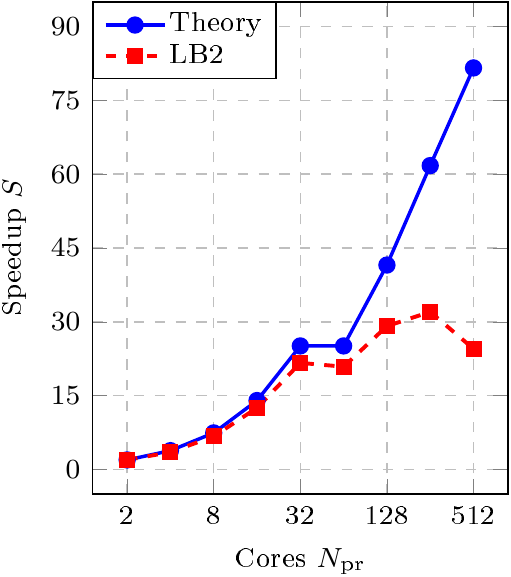}}
		\end{subfloatrow}
	}{\caption{\label{f:super_performance}Parareal's performance for the super-critical case.}}
\end{figure}

Figure~\ref{f:super_conv} depicts the Choptuik scaling that results from solutions computed with Parareal for $N_{\text{pr}}=256$ after the first three iterations. Table~\ref{t:super_values} lists the generated values of $\phi^{\star}_0$ and $\gamma$ (see Section~\ref{s:mass_scaling}), and errors compared to the value provided by the fine integrator, which agrees with the result in~\cite{Garfinkle1994a}. As can be seen in Figure~\ref{f:super_conv}, the coarse integrator $\mathcal{C}$ alone cannot adequately resolve black holes with $\phi_0-\phi^{\star}_0\lesssim10^{-9}$ (they are too small for $\mathcal{C}$ to be ``visible'') and its $\gamma$ is wrong by about $20$\%. This means that the coarse method is too ``coarse'' in the sense that, on its own, it cannot correctly capture the physics underlying the investigated problem. Nonetheless, \textit{Parareal} is not only capable of generating the correct black hole physics but can do so after only one iteration.

\begin{table}[H]
	\centering
	\begin{tabular}{llrrr}
		\toprule
			& \multicolumn{2}{c}{$\phi^{\star}_0$} & \multicolumn{2}{c}{$\gamma$} \\
			\cmidrule(lr){2-3} \cmidrule(lr){4-5}
			& \multicolumn{1}{l}{Value} & \multicolumn{1}{r}{Error (\%)} & \multicolumn{1}{l}{Value} & \multicolumn{1}{r}{Error (\%)} \\
		\midrule
			Coarse & $0.01057748$ & $7.25\cdot10^{-1}$ & $0.458$ & $20.21$ \\
			$N_{\text{it}}=1$ & $0.01055915$ & $5.51\cdot10^{-1}$ & $0.377$ & $1.05$ \\
			$N_{\text{it}}=2$ & $0.01050240$ & $1.01\cdot10^{-2}$ & $0.370$ & $2.89$ \\
			$N_{\text{it}}=3$ & $0.01050135$ & $9.52\cdot10^{-5}$ & $0.381$ & $0$ \\
			Fine & $0.01050134$ & $0$ & $0.381$ & $0$ \\
		\bottomrule
	\end{tabular}
	\caption{\label{t:super_values}Approximate values and relative errors for the critical amplitude $\phi^{\star}_0$ and resulting straight line slope $\gamma$.}
\end{table}

Figure~\ref{f:super_speed} visualizes the speedup achieved in the super-critical case including the theoretical estimate according to Equation~\eqref{e:speedup_formula}. The numbers of iterations required for Parareal to converge are derived from an analysis just like the one plotted in Figure~\ref{f:sub_conv} for the sub-critical case and basically the values are identical. Up to $64$ processes, good speedup close to the theoretical bound is observed. For larger core numbers however, speedup reaches a plateau and performance is no longer increasing. As in the sub-critical case, as $N_{\text{pr}}$ increases, the computing times per TS eventually become too small and Parareal's runtime becomes dominated by, \textit{e.g.}, communication (see Figure~\ref{f:overhead_indicators}). Even though the temporal parallelization eventually saturates, substantial acceleration of almost a factor of $30$ using $128$ cores in time is possible, corresponding to a parallel efficiency of about $23$\%.

\section{\label{s:conclusion}Conclusion}

The article assesses the performance of the parallel-in-time integration method Parareal for the numerical simulation of gravitational collapse of a massless scalar field in spherical symmetry. It gives an overview of the dynamics and physics described by the corresponding Einstein field equations and presents the employed numerical methods to solve them. Because the system is formulated and solved in characteristic coordinates, the computational spacetime domain is triangular so that later time steps carry fewer spatial degrees-of-freedom. A strategy for balancing computational \textit{cost} per subinterval instead of just number of steps is discussed and its benefits are demonstrated by traces using the \texttt{Vampir} tool. Numerical experiments are presented for both the sub- and super-critical case. Parareal converges rapidly for both and, for the latter, correctly reproduces Choptuik's mass scaling law after only one iteration despite the fact that the used coarse integrator alone generates a strongly flawed mass scaling law. This underlines the capability of Parareal to quickly correct a coarse method that does not resolve the dynamics of the problem. The results given here illustrate that Parareal and presumably other parallel-in-time methods as well can be used to improve utilization of parallel computers for numerical studies of black hole formation.

Multiple directions for future research emerge from the presented results. Evaluating performance gains for computing the critical solution~\cite{Choptuik1993a,GundlachAndMartinGarcia2007a} would be valuable. Next, complexer collapse scenarios such as in the Einstein-Yang-Mills system~\cite{Choptuik1999a}, axial symmetry~\cite{Pretorius2002a}, or binary black hole spacetimes~\cite{Pretorius2005a} could be addressed. An extended implementation of Parareal could utilize a more sophisticated convergence criterion~\cite{Aubanel2011a}, a more flexible black hole detection, and parallelism in space via, \textit{e.g.}, again Parareal. The latter would be possible because the integration along the characteristic we took to represent space is for the solution of initial value problems just like in the temporal direction. Another topic of interest is that of adaptive mesh refinement~\cite{Thornburg2015a}: how it can be used efficiently in connection with Parareal or other time parallel methods seems to be an open problem. As discussed in the introduction, a mathematical analysis of the convergence behavior of Parareal for Einstein's equations would be of great interest as well, particularly since the good performance is unexpected in view of the negative theoretical results for basic hyperbolic problems. Finally, incorporating a parallel-in-time integration method into a software library widely used for black hole or other numerical relativity simulations would be the ideal way to make this new approach available to a large group of domain scientists.\footnote{A copy of the library \texttt{Lib4PrM} for the Parareal method can be obtained by cloning the \texttt{Git} repository \href{https://scm.ti-edu.ch/projects/lib4prm/}{https://scm.ti-edu.ch/repogit/lib4prm}.}

\section*{Acknowledgments}

We would like to thank Matthew Choptuik from the University of British Columbia in Vancouver, Canada and Jonathan Thornburg from the Indiana University in Bloomington (IN), United States of America for providing feedback and suggestions on an earlier version of the manuscript. We would also like to thank Jean-Guillaume Piccinali and Gilles Fourestey from the Swiss National Supercomputing Center (CSCS) in Lugano, Switzerland and Andrea Arteaga from the Swiss Federal Institute of Technology Zurich (ETHZ), Switzerland for discussions concerning the hardware at CSCS.

This research is funded by the Deutsche Forschungsgemeinschaft (DFG) as part of the ``ExaSolvers'' project in the Priority Programme 1648 ``Software for Exascale Computing'' (SPPEXA) and by the Swiss National Science Foundation (SNSF) under the lead agency agreement as grant SNSF-145271. The research of A.K., D.R., and R.K. is also funded through the "FUtuRe SwIss Electrical InfraStructure" (FURIES) project of the Swiss Competence Centers for Energy Research (SCCER) at the Commission for Technology and Innovation (CTI) in Switzerland.

{\begingroup
\raggedright
\setlength{\bibsep}{0\baselineskip}
\bibliographystyle{abbrvnat}
\bibliography{biblio,pint}
\endgroup}

\end{document}